\documentstyle[12pt]{article}

\begin{document}
\large
\begin{center}
{\Large\bf Something about spin-fermion connection}\\
\ \\
Stanislav V. Dobrov\\
Institute of Electrophysics,\\
106 Amundsen st., Ekaterinburg 620016, Russia\\
E-mail: stas@iep.uran.ru
\end{center}
\begin{abstract}
Jordan-Wigner-type transformations connecting the spin-$\frac32$ operators and
two kinds of fermions are derived. A general condition of fermionizability
of spins is obtained and a theorem establishing connection between
half integer spins and fermions is proved. The fallibility of a previous attempt to 
generalize the Jordan-Wigner transformation for all spins
(C.D.~Batista and G.~Ortiz, Phys. Rev. Lett. {\bf 86}, 1082 (2001)) is pointed
out.
\end{abstract}

\section{Introduction}
Variables exchange very often are used during solving of different problems.
Such transformations are carried out in both classical and quantum mechanics.
Sometimes they essentially simplify the analisis of the problem under consideration.

One of the most known nontrivial variables exchange in quantum many body theory
is the Jordan-Wigner transformation (JWT) suggested by P.~Jordan and
E.~Wigner in 1928 \cite{Jordan}.
It establishs the connection between 1D lattice
spins $\frac12$ and spinless fermions on the same lattice. The most famous
applications of the JWT are solution of spin-$\frac12$
$XY$-chain by E.~Lieb, T.~Schultz and D.~Mattis \cite{Lieb} and the very effective
free fermions method developed by the same researchers
to calculate thermodynamic quantities of the 2D Ising model \cite{Schultz}.
Quite recently
this transformation was generalized for 2D \cite{Fradkin} and for D$>$2
\cite{Huerta} spin-$\frac12$ systems. 

But in nature all the spins up to $\frac{15}{2}$ exists \cite{Mila}.
Very interesting
and important to understand the behaviour of the spin higher than $\frac12$
systems. The study of quantum models with such spins demands nonstandart
methods. Can the JWT-type transformations for higher spins
help us to solve these problems ? To answer this question we
first of all must try to derive such relations between spins and fermions.
Recently the attempt to generalize JWT was undertaken by
C.D.~Batista and G.~Ortiz \cite{Batista}. Unfortunately,
their attempt was indeed unsuccessful.
Below we shall demonstrate this. Then we shall construct explicitly true
transformations between spins $\frac32$ and fermions of two kinds,
derive a connection between spin $\frac52$ and fermion and prove a general
theorem establishing connection between half integer spins and fermions.

\section{Mistake of C.D. Batista and G. Ortiz.}
We shall find the mistake of Batista and Ortiz considering two
examples. First --- for integer spin, second --- fo half-odd integer spin.

To begin we shall examine Batista and Ortiz transformations (BOT) for
the spin 1. We shall be
speaking on the language of usual fermi-operators of creation and
annihilation but not of Hubbard operators as it was in \cite{Batista}.
Also we shall be omitting in this section
the site indexes and the string operators as nonessential things
for our consideration. Then the BOT can be written as
\begin{equation}
S^+=\sqrt{2}((1-n_1)c_2^{\dag}+(1-n_2)c_1)\,,
\end{equation}
\begin{equation}
S^-=\sqrt{2}((1-n_1)c_2+(1-n_2)c_1^{\dag})\,.
\end{equation}
Here $S^\pm=S^x\pm {\rm i}S^y\,$; $S^x$, $S^y$, $S^z$ --- the operators of spin
components, and the next commutative relations must be fulfiled:
\begin{equation}
[S^+,\,S^-]=2S^z\,,\quad [S^z,\,S^\pm]=\pm S^\pm\,,\quad
 \{S^+,\,S^-\}=2S(S+1)-2(S^z)^2\,.
\end{equation}
In our case $S=1$. $c_{1,2}$ and $c_{1,2}^{\dag}$ are the annihilation and
creation operators for fermions of first (labeled as 1) and second
(labeled as 2) kinds with the next commutative relations:
\begin{equation}
\{c_{\alpha},\,c_{\beta}\}=0\,,\ 
\{c_{\alpha}^{\dag},\,c_{\beta}^{\dag}\}=0\,,\ 
\{c_{\bar\alpha},\,c_{\alpha}^{\dag}\}=0\,,\ 
\{c_{\alpha},\,c_{\alpha}^{\dag}\}=1\,\quad (\alpha,\,\beta=1,\,2),
\end{equation}
and $n_{\alpha}=c_{\alpha}^{\dag}c_{\alpha}\,$.

From (1) and (2) it follows:
\begin{equation}
S^+S^-=2(1-n_1)\,,
\end{equation}
\begin{equation}
S^-S^+=2(1-n_2)\,.
\end{equation}
Thus, from (3),
\begin{equation}
S^z=\frac12 [S^+,\,S^-]=n_2-n_1
\end{equation}
in agreement with \cite{Batista}, and
\begin{equation}
(S^z)^2=n_1+n_2-2n_1n_2\,.
\end{equation}
While (5) and (6) result in
\begin{equation}
\{S^+,\,S^-\}=4-2n_1-2n_2\,,
\end{equation}
the equations (3) and (8) lead us to
\begin{equation}
\{S^+,\,S^-\}=4-2n_1-2n_2+4n_1n_2\,.
\end{equation}
The last addendum in the right part of (10) absents in the right part of (9).
That is the equations (9) and (10) contradict each other. Thus the BOT for an
integer spin do not keep the spin operators algebra and hence are unacceptable.

Now we turn to BOT for half-odd integer spins. We shall consider the case of
spin $\frac32$, for which in \cite{Batista} was suggested
to introduce three kinds of flavourless fermions. According to BOT
\begin{equation}
S^+=\sqrt{3}c_1^{\dag}(1-n_2)(1-n_3)+2c_2^{\dag}c_1(1-n_3)+
\sqrt{3}c_3^{\dag}c_2(1-n_1)
\end{equation}
and $S^-=(S^+)^{\dag}$.  Then
\begin{eqnarray}
S^+S^-=3n_1(1-n_2)(1-n_3)+4n_2(1-n_1)(1-n_3)+\nonumber\\
+3n_3(1-n_2)(1-n_1)\,,
\end{eqnarray}
\begin{eqnarray}
S^-S^+=3(1-n_1)(1-n_2)(1-n_3)+4n_1(1-n_2)(1-n_3)+\nonumber\\
+3n_2(1-n_1)(1-n_3)\,.
\end{eqnarray}
So
\begin{equation}
\{S^+,S^-\}=3+4n_1+4n_2-11n_1n_2-7n_1n_3-7n_2n_3+14n_1n_2n_3\,,
\end{equation}
but
\begin{eqnarray}
(S^z)^2=\frac14 ([S^+,S^-])^2=
\frac94-2n_1-2n_2+\frac{7}{4}n_1n_2-\nonumber\\
-\frac{1}{4}n_1n_3-\frac14 n_2n_3+\frac{1}{2}n_1n_2n_3\,,
\end{eqnarray}
and we see that the third relation from (3) is not satisfied. Consequently, BOT
for half-odd integer spins are wrong too.

The general conclusion is that the BOT are erroneous.

In the next section we shall construct true transformations between
spin-$\frac32$ operators and two kinds of fermions.
From procedure which will be used it will be clear what spins can be
represented through fermions.

\section{Fermionization of spin $\frac32$}
Firstly we shall consider one-spin problem and after its solution generalize
on a lattice.

Basis of the spin state space consists of four vectors $|-3/2\rangle$,
$|-1/2\rangle$, $|1/2\rangle$, $|3/2\rangle$ being the eigenvectors of $S^z$
operator. Phase space of two kinds fermions is four-dimensional
too with the basis $|0\rangle$, $|1\rangle=c_1^{\dag}|0\rangle$,
$|2\rangle=c_2^{\dag}|0\rangle$, $|1\&2\rangle=c_2^{\dag}c_1^{\dag}|0\rangle$.
Now we need establish a connection between spin operators and operators acting
in fermionic space. Let write out all the needed matrix representations of
these operators:
\begin{equation}
S^+=
\left(
\begin{array}{cccc}
0&\sqrt{3}&0&0\\
0&0&2&0\\
0&0&0&\sqrt{3}\\
0&0&0&0
\end{array}
\right),\quad
S^z=\left(
\begin{array}{cccc}
\frac32&0&0&0\\
0&\frac12&0&0\\
0&0&-\frac12&0\\
0&0&0&-\frac32
\end{array}
\right),
\end{equation}
and $S^-=(S^+)^{\dag}=(S^+)^{\rm T}$;
\begin{equation}
c_1=
\left(
\begin{array}{cccc}
0&0&0&0\\
0&0&0&0\\
-1&0&0&0\\
0&1&0&0
\end{array}
\right),\quad
c_2=\left(
\begin{array}{cccc}
0&0&0&0\\
1&0&0&0\\
0&0&0&0\\
0&0&1&0
\end{array}
\right),
\end{equation}
and $c_{1,2}^{\dag}=(c_{1,2})^{\dag}=(c_{1,2})^{\rm T}$. This matrices are
sufficiently simple and one can easily see that
the solution of the one-spin problem is:
\begin{equation}
\left\{
\begin{array}{l}
S^-=\sqrt{3}c_2+2c_2^{\dag}c_1\\
S^+=\sqrt{3}c_2^{\dag}+2c_1^{\dag}c_2
\end{array}
\right.\,.
\end{equation}
It is easily to check that all the relations (3) with $S=\frac32$ are
satisfied by (18) with
\begin{equation}
S^z=\frac12 [S^+,\,S^-]=-\frac32+2n_1+n_2\,.
\end{equation}

By the passage to a lattice we must introduce the commutativity of all spin
operators acting on different sites. Hence we must modify (18):
\begin{equation}
\left\{
\begin{array}{l}
S_i^-=\sqrt{3}c_{2i}U_i+2c_{2i}^{\dag}c_{1i}\\[\bigskipamount]
S_i^+=\sqrt{3}U_i^{\dag}c_{2i}^{\dag}+2c_{1i}^{\dag}c_{2i}
\end{array}
\right.\,,
\end{equation}
where for 1D lattice
$U_i=\exp\left\{{\rm i}\pi\sum\limits_{j<i}(n_{1j}+n_{2j})\right\}$
is the string operator. The form (19) for $S_i^z$ with site index
is not changed.

The inverse transformations are obtained in the same way. And for a lattice
we have
\begin{equation}
\left\{
\begin{array}{ll}
c_{1i}=-\frac{1}{\sqrt{3}}S_i^-S_i^zS_i^-W_i\,,&
c_{1i}^{\dag}=-\frac{1}{\sqrt{3}}W_i^{\dag}S_i^+S_i^zS_i^+\,,
\\[\bigskipamount]
c_{2i}=\frac{1}{\sqrt{3}}\left(\frac12+S_i^z\right)^2S_i^-W_i\,,&
c_{2i}^{\dag}=\frac{1}{\sqrt{3}}W_i^{\dag}S_i^+\left(\frac12+S_i^z\right)^2\,,
\end{array}
\right.
\end{equation}
where for 1D lattice $W_i=\prod\limits_{j<i}X_j$ with
\begin{equation}
X_j=\frac54-(S_j^z)^2=
\exp\left\{{\rm i}\frac{\pi}{2}\left(\frac14-(S_j^z)^2\right)\right\}\,.
\end{equation}
All the relations (4) for one site are fulfiled and the operator $W_i$ ensures
anticommutativity the $c$-operators acting on different sites. The form
of $X_j$ one can construct from requirement
\begin{equation}
X_jc_{\alpha j}=-c_{\alpha j}X_j\,.
\end{equation}
Assuming that $X_j$ has a form
$$
X_j=1+\xi S_j^z+\zeta (S_j^z)^2+\eta (S_j^z)^3\,
$$
with parameters $\xi$, $\zeta$ and $\eta$ which must be defined from (23),
taking into account (19) and introducing the normalization factor $\frac54$ we
shall get (22). Note that $X_j=X_j^{\dag}$ and $X_j^2=1$.

Obtained transformations (19), (20) and (21) are sufficiently complicated
and much efforts will be needed to derive some profit from them.
Let try to transform the isotropic $XY$ spin-$\frac32$ chain in fermionic
form using (20):
\begin{eqnarray}
H=J\sum\limits_i (S_i^x S_{i+1}^x+S_i^y S_{i+1}^y)=
\frac12 J\sum\limits_i (S_i^+ S_{i+1}^- +S_i^- S_{i+1}^+)=
\nonumber\\[\bigskipamount]
=\frac12 J\sum\limits_i \left(3c_{2i}^{\dag}(1-2n_{1i})c_{2i+1}+
2\sqrt{3}U_i(c_{1i+1}^{\dag}c_{2i+1}c_{2i}-c_{1i}^{\dag}c_{2i}c_{2i+1})+
\right.
\nonumber\\[\bigskipamount]
\left.+4c_{2i+1}^{\dag}c_{1i+1}c_{1i}^{\dag}c_{2i}+\mbox{h.c.}\right)\,.
\end{eqnarray}

We do not see the free fermions Hamiltonian as it was for spin-$\frac12$ case
\cite{Lieb}. Now fermions interact with the interaction being very
complex due to operator
$U_i$ (string operator in 1D case).

Note that usual JWT one can get using the
procedure described in this section.

So, we have explicitly constructed Jordan-Wigner-type transformations between
spin-$\frac32$ and two kinds of flavourless fermions. This transformations
can be used to map spin-$\frac32$ systems onto fermionic models with two kinds
of fermions: with spin up and with spin down. And vice versa --- to map
fermionic model onto spin model.

Further the next questions arise. What can one say
about spin 1 fermionization ? What spins can we fermionize ? Procedure
applied above to fermionize the spin $\frac32$ prompt us the answer: only
operators of spins $S$, satisfying the equality
\begin{equation}
2S+1=2^n\,,\quad n\in{\rm N}
\end{equation}
can be expressed in terms of fermions. The matrices of spin-$S$ operators
being the infinitesimal matrices of irreducible representation
of $SU(2)$ group with eigenvalue of Casimir operator ${\rm\bf S}^2$ being equal
$S(S+1)$ have only the size $(2S+1)\times (2S+1)$ and, of course,
can not be presented
through the matrices of other sizes. The phase space of $n$ kinds of
flavourless fermions is $2^n$-dimensional. The representation of the algebra
of fermionic operators in this space is irreducible too.
Hence the condition (25) follows.

Spin 1 do not satisfy the condition (25). So, its operators can not
be expressed in terms of fermions. The next spin suitable for fermionization
is $\frac72$ with three kinds of fermions being needed,
then spin $\frac{15}{2}$ follows with four kinds of fermions etc.
Of course, spin $\frac12$ satisfy the condition (25) with $n=1$ and
usual Jordan-Wigner transformations exist \cite{Jordan}.

\section{General result}
In previous section we have found that only the operators
of spins $S$ satisfying the equality
\begin{equation}
2S+1=2^n\,,
\end{equation}
can be expressed through only fermions. Other spins can not
be subjected to complete fermionization. However it inevitably comes to mind
that all half integer spins have to be connected with fermions. And indeed
such a connection exists. Here we shall prove a general theorem establishing
this connection. But at first we shall explicitly demonstrate it
on example of spin $\frac52$. We shall carry out a partial fermionization of this spin.
Firstly we shall get one-spin expression and then generalize on
a lattice to construct the Jordan-Wigner-type transformation \cite{Jordan}.

To begin we shall write out the matrices of spin $\frac52$:
\begin{equation}
S^+=\left(
\begin{array}{cccccc}
0&\sqrt{5}&0&0&0&0\\
0&0&2\sqrt{2}&0&0&0\\
0&0&0&3&0&0\\
0&0&0&0&2\sqrt{2}&0\\
0&0&0&0&0&\sqrt{5}\\
0&0&0&0&0&0
\end{array}
\right)\,,\quad
S^z=\left(
\begin{array}{cccccc}
\frac52&0&0&0&0&0\\
0&\frac32&0&0&0&0\\
0&0&\frac12&0&0&0\\
0&0&0&-\frac12&0&0\\
0&0&0&0&-\frac32&0\\
0&0&0&0&0&-\frac52
\end{array}
\right)
\end{equation}
and $S^-=(S^+)^{\rm T}$. The operators satisfy the commutative
relations (3).

Now let consider a system consisting of one spinless fermion and
one spin 1. The operators of the former we shall denote by $c$ and $c^{\dag}$
with the next commutative relation: $\{c,c^{\dag}\}=1$, --- and $c^2=(c^{\dag})^2=0$;
the operators of the latter we shall denote by $T^{\pm}$ and $T^z$ with
commutative relations being the same as (3) where $S$ must be replaced
by $T$.
Again we shall write out the matrices of the operators being needed further:
\begin{equation}
c^{\dag}=\left(
\begin{array}{ccc}
1&0&0\\
0&1&0\\
0&0&1
\end{array}
\right)\otimes\left(
\begin{array}{cc}
0&1\\
0&0
\end{array}
\right)=\left(
\begin{array}{cccccc}
0&1&0&0&0&0\\
0&0&0&0&0&0\\
0&0&0&1&0&0\\
0&0&0&0&0&0\\
0&0&0&0&0&1\\
0&0&0&0&0&0
\end{array}
\right)\,,
\end{equation}

\begin{equation}
T^+c=\left(
\begin{array}{ccc}
0&\sqrt{2}&0\\
0&0&\sqrt{2}\\
0&0&0
\end{array}
\right)\otimes\left(
\begin{array}{cc}
0&0\\
1&0
\end{array}
\right)=\left(
\begin{array}{cccccc}
0&0&0&0&0&0\\
0&0&\sqrt{2}&0&0&0\\
0&0&0&0&0&0\\
0&0&0&0&\sqrt{2}&0\\
0&0&0&0&0&0\\
0&0&0&0&0&0
\end{array}
\right)\,,
\end{equation}

\begin{equation}
(T^z)^2c^{\dag}=\left(
\begin{array}{ccc}
1&0&0\\
0&0&0\\
0&0&1
\end{array}
\right)\otimes\left(
\begin{array}{cc}
0&1\\
0&0
\end{array}
\right)=\left(
\begin{array}{cccccc}
0&1&0&0&0&0\\
0&0&0&0&0&0\\
0&0&0&0&0&0\\
0&0&0&0&0&0\\
0&0&0&0&0&1\\
0&0&0&0&0&0
\end{array}
\right)
\end{equation}
and we see from (27)--(30) that
\begin{equation}
\left\{
\begin{array}{l}
S^+=3c^{\dag}+2T^+c+(\sqrt{5}-3)(T^z)^2 c^{\dag}\,,
\\[\bigskipamount]
S^-=3c+2T^-c^{\dag}+(\sqrt{5}-3)(T^z)^2 c
\end{array}
\right.
\end{equation}
and
\begin{equation}
S^z=\frac12 [S^+,S^-]=-\frac12+c^{\dag}c+2T^z\,.
\end{equation}
All the relations (3) are satisfied by (31) and (32).

By the passage to a lattice we must modify (31):
\begin{equation}
\left\{
\begin{array}{l}
S_i^+=\left(3c_i^{\dag}+2T_i^+c_i+(\sqrt{5}-3)(T_i^z)^2 c_i^{\dag}\right)U_i\,,
\\[\bigskipamount]
S_i^-=U_i^{\dag}\left(3c_i+2T_i^-c_i^{\dag}+(\sqrt{5}-3)(T_i^z)^2 c_i\right)\,.
\end{array}
\right.
\end{equation}
Here the operator $U_i$ ensures commutativity of spin-$\frac52$ operators
acting on different sites and for 1D case it is the usual string operator:
\begin{equation}
U_i=\exp\left\{{\rm i}\pi\sum\limits_{j<i}c_j^{\dag}c_j\right\}\,.
\end{equation}

So, we have expressed spin $\frac52$ through one spinless fermion and one
spin 1. Being supported by this result and results obtained in previous
section,
one can maintain that each half integer spin
can be represented through either only fermions or fermions and some
integer spin. In other words from each half integer spin one can extract
fermionic degrees of freedom so that only integer spin, including spin 0,
remain. Let give a rigorous proof.

Consider half integer spin $S$. Then $2S+1$ is an even number.
Let
\begin{equation}
\left\{
\begin{array}{l}
2S+1=0\,{\rm mod}\, 2^n\\
2S+1>0\,{\rm mod}\, 2^{n+1}
\end{array}
\right.\ ,\quad \mbox{for some } n\in {\rm N}
\end{equation}
Further consider an {\it integer} spin $T$ such as
\begin{equation}
2T+1=\frac{2S+1}{2^n}
\end{equation}
The matrices of the spin $T$ operators algebra representation that is
the matrices of the form $(T^+)^k (T^z)^l$ and $(T^z)^l(T^-)^k$
$(k,\ l=0,\ldots,2T)$ originate
the complete (and even overcomplete) set in the space of matrices
of the size $(2T+1)\times(2T+1)$. Denote these matrices as $A_i$
$(i=1,\ldots, (2T+1)^2)$. The matrices of $n$ flavourless fermions
operators representation that is the matrices of the form
$(c_1^{\dag})^{k_1}\ldots (c_n^{\dag})^{k_{n}}(c_n)^{l_{n}}\ldots(c_1)^{l_1}$
$(k_{\alpha},\ l_{\beta}=0,\ 1;\ \alpha,\ \beta=1,\ldots,n)$ 
in the phase space of these fermions
originate the complete set in the space of matrices of size $2^n\times 2^n$.
Denote these matrices as $\mit\Gamma_j$ $(j=1,\ldots, 2^{2n})$.
Then the set of matrices $A_i\otimes\mit\Gamma_j$ originate the complete set
in the space of matrices of size
$\left[(2T+1)2^n\right]\times\left[(2T+1)2^n\right]=(2S+1)\times(2S+1)$.
Hence the next decompositions exist
\begin{equation}
S^{\pm}=\sum\limits_{i=1}^{(2T+1)^{2}}\sum\limits_{j=1}^{2^{2n}}
\alpha_{ij}^{\pm}A_i\otimes\mit\Gamma_j\,,\quad
S^z=\sum\limits_{i=1}^{(2T+1)^2}\sum\limits_{j=1}^{2^{2n}}
\alpha^z_{ij}A_i\otimes\mit\Gamma_j\,,
\end{equation}
Q.E.D. Here $\alpha_{ij}^{\pm,\,z}$ are some numbers.
Note that simultaneously we once again derive the condition of
complete fermionization: $2S+1=2^n$, --- when $T=0$. 

By the passage to a lattice all the fermion-like operators
have to get
the same factor $U_i$ in $S_i^+$ or $U_i^{\dag}$ in $S_i^-$
which in 1D case is the usual string operator:
\begin{equation}
U_i=\exp\left\{{\rm i}\pi\sum\limits_{j<i} \sum\limits_{\alpha=1}^{n}
c_{\alpha j}^{\dag}c_{\alpha j}\right\}\,.
\end{equation}
Here $i$ and $j$ are the site indexes.

Unfortunetly, further fermionization that is a complete fermionization of
integer
spins is impossible. Let adduce another argument, being additional to the
argument produced above, in favor of this
statement. The first term in the right side of the third relation in (3)
inevitably make us to deal with the full fermionic algebra being considered
in trying to fermionize a spin $S$.
Number of basis elements in each fermionic algebra is an even number.
But the number of basis elements in the algebra of spin $S$ is equal to $(2S+1)^2$
and for an integer spin is an odd number. It is clear that there is no
isomorphism between these algebras.

\section{Conclusion}
Thus we have shown that each half integer spin $S$ such as
the equality and inequality
(35) fulfil can be represented in terms of $n$ fermions with $n$
being the same as in (35) and some integer spin, including spin 0.
Obtained result can be applied in investigations of different quantum
spin systems.

\end{document}